# An Architecture Process Maturity Model of Software Product Line Engineering

**Faheem Ahmed, Luiz Fernando Capretz**

**Department of Electrical & Computer Engineering, Faculty of Engineering
University of Western Ontario, London, Ontario, Canada, N6A 5B9**

**f.ahmed@uaeu.ac.ae, lcapretz@eng.uwo.ca**

**Abstract:** Software architecture has been a key research area in the software engineering community due to its significant role in creating high quality software. The trend of developing product lines rather than single products has made the software product line a viable option in the industry. Software product line architecture is regarded as one of the crucial components in the product lines, since all of the resulting products share this common architecture. The increased popularity of software product lines demands a process maturity evaluation methodology. Consequently, this paper presents an architecture process maturity model for software product line engineering to evaluate the current maturity of the product line architecture development process in an organization. Assessment questionnaires and a rating methodology comprise the framework of this model. The objective of the questionnaires is to collect information about the software product line architecture development process. Thus, in general this work contributes towards the establishment of a comprehensive and unified strategy for the process maturity evaluation of software product line engineering. Furthermore, we conducted two case studies and reported the assessment results, which show the maturity of the architecture development process in two organizations.

## I. INTRODUCTION

Recently, software development trends have caused single product development to evolve into "software product line architecture" (SPLA), which integrates lines of resulting products. The main objective of SPLA is to reuse the architecture for successive product development. Clements [7] defines the term "software product line" (SPL) as a set of software intensive systems sharing a common, managed set of features that satisfy the specific needs of a particular market segment and are developed from a common set of core assets in a prescribed way. The SPL is receiving an increasing amount of attention from software development organizations because of the promising results in cost reduction, quality improvements and reduced delivery time. Clement et al. [8] report that SPL engineering is a growing software engineering sub-discipline and many organizations, including Philips[®], Hewlett-Packard[®], Nokia[®], Raytheon[®], and Cummins[®], are using it to achieve extraordinary gains in productivity, development time, and product quality. European researchers present many other corresponding terminologies for the SPL such as "product family", "product population" and "system family". The architecture dimension of the SPL concept has interested many researchers, and the architectural aspects of the SPL, such as domain engineering, product line architecture, and commonality and variability management, have been a key area of research since the introduction of the concept in the mid-nineties. For the last decade, research has been conducted on the SPL process methodology including product line architecture, commonality and variability management, core assets management, business case engineering, and application and domain engineering [6][9][22][42].

SPL engineering is gaining popularity in the software industry. Some of the potential benefits of this approach include cost reduction, improvement in quality and a decrease in product development time. The increasing popularity of SPL engineering necessitates a process maturity evaluation methodology. To date, no work has been reported in this area apart from a few initial theoretical studies. Accordingly, this study presents an Architecture Process Maturity Model (APMM) of SPL engineering for evaluating the maturity of an organization's product line architecture development process. The framework of this model assesses the maturity of the SPLA development process according to the way in which sets of various architecture development activities are aligned with the SPL engineering methodology. In particular, assessment questionnaires and a rating methodology comprise the framework of this model. The objective of the questionnaires is to collect information about the SPLA development process. We applied the model to two organizations and demonstrate the assessment result in subsequent sections of this paper. Apart from its general and specific limitations, the APMM presented in this paper contributes significantly to the area of SPL by addressing a topic of immense importance.

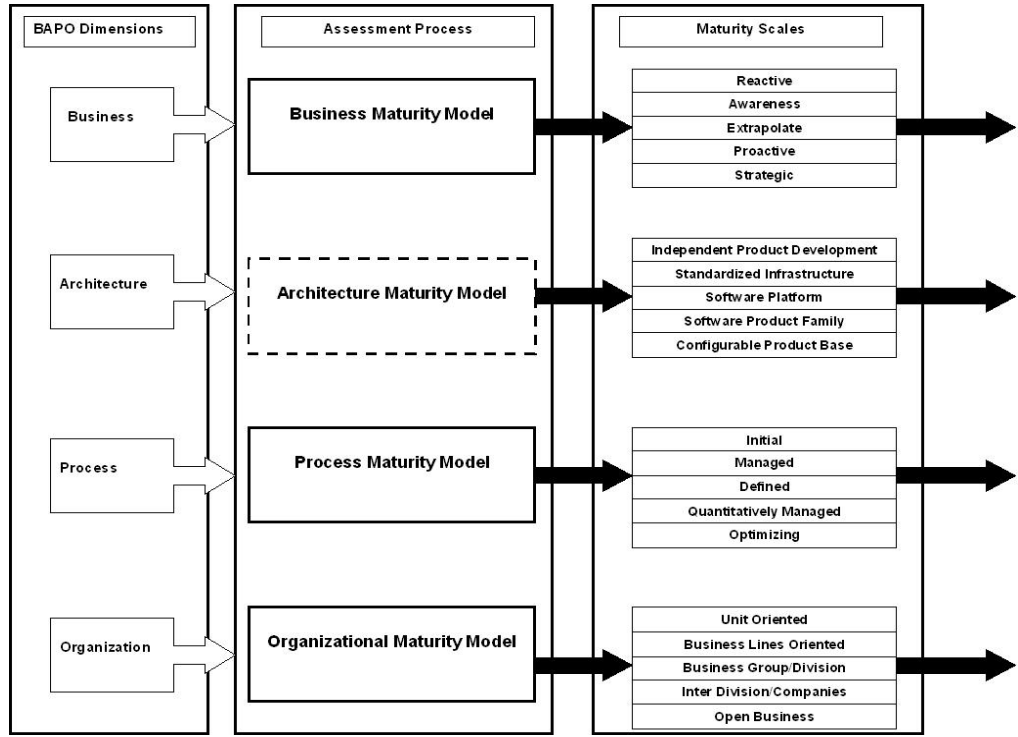

Figure 1: Software Product Line Engineering Maturity Model: The Big Picture (Based on [38])

## II. SOFTWARE PRODUCT LINE ENGINEERING MATURITY MODEL: THE BIG PICTURE

The software product line is a relatively new concept in the history of software development and business. A lot of effort has been spent on the process methodology and the industrialization of this paradigm. Software product line process assessment is a relatively new area of research in which, so far, very little work has been done. Currently, researchers from both academia and industry are attempting to develop a prescribed and systematic way of measuring the maturity of a software product line process. Jones and Soule [23] discuss the relationships between the software product line process and the CMMI, and observe that the software engineering process discipline as specified in the CMMI provides an essential foundation for the software product line process. These researchers conclude that apart from the key process areas of the CMMI model, the software product line requires the mastery of many other essential practice areas. Although Jones and Soule have compare the key process areas of the software product line with the CMMI-model and find some similarities, they do not discuss any procedure to evaluate the maturity of the software product family process. Moreover, they state that there is a need to establish a comprehensive strategy for the process assessment of the software product line in particular, which is what this research aims to accomplish. As previously mentioned, the SEI proposed the Product Line Technical Probe (PLTP) [9], which is aimed at analyzing an organization's ability to adapt and succeed with the software product line approach. The framework of PLTP is divided into three categories of product development, core assets development, and management. However, the framework does not clearly define any maturity levels and the procedure to evaluate maturity of the software product line process. Rather it identifies potential areas of concern that require attention while carrying out that software product line process, and it also presents a framework to set up a software product line within an organization.

The acronym BAPO [38] (Business-Architecture-Process-Organization) defines process concerns associated with the SPL. The dimensions of business, architecture, process and organization are considered critical because they establish an infrastructure and manage the profitability of the products resulting from a SPL. Specifically, the architectural dimension of BAPO is important because it deals with the technical means to build a framework that will be shared by a number of products from the same family. *van der* Linden et al. [38] propose a four-dimensional SPL maturity evaluation framework based on the BAPO concept of operations. It provides an early foundation for a systematic and comprehensive strategy to perform a process maturity evaluation of SPL. Figure 1 illustrates the conceptual layout of this maturity evaluation method. As previously discussed, the four dimensions of the framework are labeled as Business, Architecture, Process and Organization. According to this conceptual layout, the overall maturity assessment of the SPL engineering comprises four separate maturity assessment models for each of the BAPO dimensions. The maturity models for each dimension of business, architecture, process and organization have not yet been given much attention by the software engineering community. *van der* Linden et al. [38] identify maturity scales of up to five levels in ascending order for each dimension of BAPO, shown in the column of "Maturity Scales" in Figure 1. In the case of a SPL, this results in separate values for each of the four dimensions.

Although *van der* Linden et al. [38] illustrates the conceptual layout of the comprehensive maturity assessment of software product line engineering and a framework to evaluate each dimension but did not provide the mechanism of assessing maturity of each dimension of software product line engineering process such as questionnaires and rating methodology which are the core features of software engineering process assessment approaches. The major contribution and further enhancement in the work of *van der* Linden et al. [38] as presented in this paper is an APMM of SPL consisting of set of questionnaires and rating methodology, thus addressing one of the critical dimensions in the SPL engineering process. The model provides a methodology for evaluating the current maturity of the SPLA development process in an organization. The maturity models that evaluate the other three dimensions of BAPO are beyond the scope of this study since this work concentrates only on the architecture dimension. Accordingly, the dotted rectangle in Figure 1 clearly highlights the scope of the work presented in this paper in the domain of process maturity assessment of SPL engineering. Thus, the main objective of this research is to contribute towards a unified strategy for the process evaluation of SPL engineering.

## A. Architecture Dimension of Software Product Line: Literature Review

The literature survey of related SPLA studies exposes some key architecture process activities such as domain engineering, commonality and variability management, requirements modeling, architecture documentation and architecture evaluation, all of which are currently in practice. We used these architecture process activities in developing the SPL maturity model presented in this paper. This section provides theoretical information and a detail discussion of these key architecture process activities in context of work carried out in the domain of SPL engineering.

Software architecture has a history of evolution. Over the last decade, the software industry has been observing and reporting modifications and advancements in technology. According to Garlan and Perry [18] software architecture includes the structure of the components of a program or system, their interrelationships, and the principles and guidelines governing their design and evolution. In this modern era, software architecture is being restructured towards a SPLA, where the focus is not single product development but rather on multiple product development. In a SPLA, all of the products share the same architecture. Pronk [35] defines SPLA as a system of reuse in which the same software is recycled for an entire class of products, with only minimal variations to support the diversity of individual product family members. According to Jazayeri et al. [22] SPLA defines the concepts, structures and textures necessary to achieve variation in the features of diverse products while ensuring that the products share the maximum amount of parts in the implementation. Mika and Tommi [32] explain that SPLA can be produced in three different ways: from scratch, from an existing product group or from a single product. Hence, software product line architecture is an effective way to minimize risks and to take advantage of opportunities such as complex customer requirements, business constraints and technology. Nevertheless, the success of SPLA depends on more than technical excellence [13]. *van der* Linden et al. [38] identify some main factors for evaluating the architecture dimension of SPL. These factors include software product family architecture, product quality, reuse levels and software variability management. Furthermore, these authors classified the architectural maturity of the SPL into five levels. In ascending order, these levels include "independent product development", "standardized infrastructure", "software platform", "software product family" and "configurable product base".

Bayer et al. [4], at the Fraunhofer Institute of Experimental Software Engineering (IESE), developed a methodology called PuLSE (Product Line Software Engineering) for the purpose of enabling the conception and deployment of SPL within a large variety of enterprise contexts. As part of the PuLSE methodology, PuLSE-DSSA develops the reference architecture for a SPL. Knauber et al. [26] explain that the basic idea of PuLSE-DSSA is to incrementally develop reference architecture guided by generic scenarios that are applied in decreasing order of architectural significance. Researchers at Philips® developed the Component-Oriented Platform Architecting (CoPAM) [2] method for the SPL of electronics products. CoPAM assumes a strong correlation among facts, stakeholder expectations, and existing architecture. Weiss and Lai [42] discuss the development of the Family-Oriented Abstraction Specification and Translation (FAST) method for the SPL process and its successful use it at Lucent Technologies®. The FAST method entails a full SPL engineering process with specific activities and targeted objects. It divides the overall process of the SPL into three major steps of domain qualification, domain engineering and application engineering. Similarly, researchers at IESE developed a methodology called KobrA [3], which defines the objects and activities involved in the SPL the engineering process. In KobrA, the process of SPL engineering is divided in to framework engineering and application engineering, and then it is further classified into sub-steps of both engineering types. These steps cover the implementation, release, inspection and testing aspects of the product line engineering process. Kang et al. [24] propose a Feature Oriented Reuse Method (FORM) to detail the aspects of SPL; FORM is an extension of the Feature-Oriented Domain Analysis (FODA) method. Also, FORM provides a methodology for using feature models in developing domain architectures and reusable components.

Although the concepts of commonality and variability management belong to domain engineering, they have been increasing in popularity over time due to their extensive use in SPLA. According to Coplien et al. [11] commonality and variability analysis gives software engineers a systematic way of conceptualizing and identifying the product family that they are creating. Kang et al. [24] discuss the use of feature models to manage commonality and variability in SPL. Furthermore, Lam [28] presents variability templates and a hierarchy-based variability management process. Thompson and Heimdah [36] propose a set-based approach to structure commonalities and variability in SPLs, whereas Kim and Park [25] describe the goal and scenario driven approach for managing commonality and variability. Ommering [39] observes that the commonalities are embodied in the overall architecture of a SPL, while the differences result from specifying variation points. Researchers [26] [31] [42] stress that the SPLA must address variability and commonality in product development. Birk et al. [5] stress that an organization dealing with SPLA should describe the architecture using well-established notations such as the UML, and the architectural description should cover all relevant views and use clearly defined semantics. Gomma and Shin [19] describe a multiple-view meta-modeling approach for SPLs using the UML notation. Zuo et al. [44] present the use of problem frames for product line engineering modeling and requirements analysis. Dobrica and Niemelä [13] discuss how UML standard concepts can be extended to address the challenges of variability management in SPLA. Eriksson et al. [15] describe a product line use case approach named PLUSS (Product Line Use case modeling for Systems and Software engineering). Etxeberria and Sagardui [16] highlight the issues that arise when evaluating product line architecture as opposed to single system architecture. More specifically, Graaf et al [20] present a scenario-based SPL evaluation technique, which provides guidelines for applying a scenario-based assessment to a SPL context by using the qualitative technique of software architecture evaluation. *van der* Hoek et al [37] propose service utilization metrics to assess the quality attributes of SPLA. Similarly, Zhang et al. [43] study the impact of variants on quality attributes using a Bayesian Belief Network (BBN), and they design a methodology applicable to SPLA evaluation. De Lange and Kang [12] propose a product-line architecture prototyping approach that uses a network technique to assess issues related to SPLA evaluation. Furthermore, Gannod and Lutz [17] define an approach to evaluate the quality and functional requirements of SPLA. Niemelä et al [33] discuss the basic issues of product family architecture development and present an evaluation model of the software product family in an industrial setting.

## III. AN ARCHITECTURE PROCESS MATURITY MODEL OF SOFTWARE PRODUCT LINE ENGINEERING

The APMM of SPL engineering aims at establishing a comprehensive strategy to evaluate the architecture dimension of the SPL process. It describes the SPLA process assessment methodology and determines the current maturity of the SPLA development process in an organization. Furthermore, it is structured to determine how various architecture process activities are conducted in SPL development. The maturity assessment of the SPLA development process assumes a strong degree of coordination between product line engineering and architecture-related process activities. It evaluates the maturity of the SPLA development process as a function of how various architecture process activities are aligned with product line engineering. The model's functional structure consists of a set of questionnaires purposely designed for evaluating the maturity at each of the five levels. A survey of work carried out in the SPLA provides foundations for designing the questionnaires, which are divided into sets of various key architecture process activities.

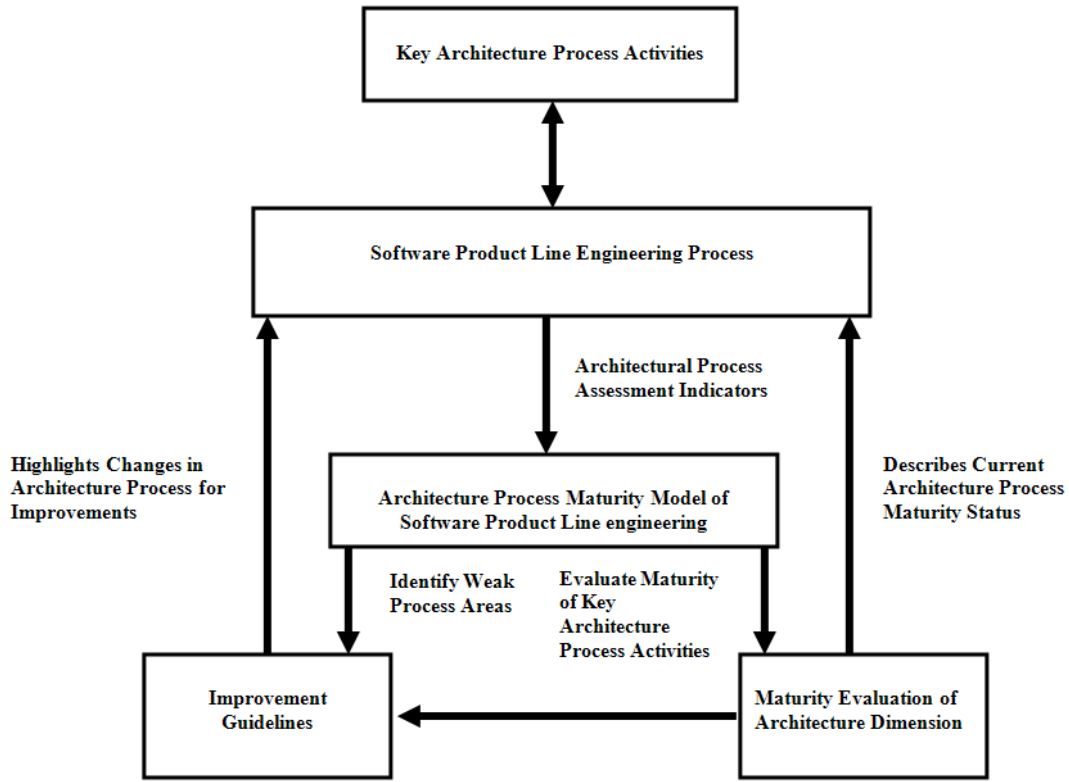

**Figure 2: Scope of Architecture Process Maturity Model of Software Product Line Engineering**

## A. General Scope of Architecture Process Maturity Model

The assessment of architecture dimension of software product line engineering is an essential activity for improving the overall software product line engineering process in an organization. The general objective of a maturity assessment model in software engineering has two folds. First, it provides mechanism to perform assessment and secondly it provides further guidelines to introduce changes in the current process to make improvements. The software product line engineering process like others requires improvements over time. However, it is very difficult to develop an efficient and effective improvements plan unless it is based on the results of a comprehensive assessment exercise. Figure 2 illustrates the framework of a comprehensive architecture process assessment exercise for the organizations dealing with software product line development. The overall software product line engineering process involves many key architecture process activities. They are blended into the software product line engineering process and can be used as indicators for the process assessment of architecture dimension of software product line engineering. For example if we take an instance of product line engineering process such as variability management, then this task requires a comprehensive domain engineering and requirements management activities. Which narrates that these key architecture process activities further facilitates the product line engineering process. An improved architecture process activity further helps in executing product line engineering process. The architecture process maturity model presented in this work uses the key architecture process activities in developing a comprehensive framework consists of questionnaires and maturity levels to carry out assessment. Furthermore, architecture process maturity assessment determines the current status of the architectural dimension of the software product line engineering process in an organization. The assessment process yields a number of recommendations based on the identification of weaknesses in the current process for improvement. Ideally after the assessment process the improvement guidelines highlights changes in current software product line engineering process to introduce improvements based on the assessment activities. However, the maturity model presented in this work does not provide any guidelines for the improvement process, which we consider as a future project for this study.

## B. Configuration of Architecture Process Maturity Model

The functional configuration of the APMM of a SPL consists of six key architecture process activities. Specifically,

Table-I defines the hierarchy and domains of the APMM of a SPL. In this paper, we use the term "architecture process activities" to refer to the practice, which, in conjunction with SPL engineering, contributes to the development and management of the SPLA. The six key architecture process activities used in this model include domain engineering, requirements management and modeling, commonality management, variability management, architecture analysis and evaluation, and architecture artifact management. We divided these six architecture process activities into a set of three dimensions that included "architecture design", "product line management" and "documentation". In comparison, Hofmeister et al. [21] classify architecture analysis, architecture synthesis and architecture evaluation as key activities performed during the architecture designing phase. The dimension of "architecture design" covers domain engineering, requirements management and modeling, architecture analysis and evaluation activities. The "product line management" dimension covers the commonality and variability management of product line architecture. Ahmed and Capretz [1] conducted an empirical investigation and found that these six architecture process activities have positive impact on the performance of software product line development in an organization. This empirical investigation motivates the inclusion of these six architecture process activities into this maturity model development. These six, significantly important architecture process activities comprise the foundation of the questionnaires, which include "statements" regarding the effectiveness of the activities as they contribute to the development and management of the SPLA.

**Table I: Configuration of Architecture Process Maturity Model**

| Dimension | Activity No. | Key Architecture Process Activities |
|---|---|---|
| Architecture Design | 1 | Domain Engineering |
|  | 2 | Requirements Management and Modeling |
|  | 3 | Architecture Analysis and Evaluation |
| Product Line Management | 4 | Commonality Management |
|  | 5 | Variability Management |
| Documentation | 6 | Architecture Artifact Management |

## C. Framework of Architecture Process Maturity Model

The concept of ranking is important for defining maturity levels in software process assessment methodologies. Many popular software process assessment frameworks, such as CMM [34] and BOOTSTRAP [27], utilize the process of ranking in defining maturity levels. Our proposed APMM also uses the approach of ranked maturity. *van der* Linden et al. [38] defines five maturity scales for the architecture dimension of a SPL. In ascending order, these levels include "independent product development", "standardized infrastructure", "software platform", "software product family" and "configurable product base". Accordingly, this study uses *van der* Linden's architecture maturity scale to develop a framework consisting of a set of questionnaires for each maturity level. Each questionnaire contains a number of statements that are divided into six key architecture process activities. The maturity level of each organization's SPLA process is determined by the extent to which the organization agrees with each statement in the questionnaire. All questionnaires shown in this paper are designed and written specifically for this APMM of software product line engineering.

Table II illustrates the framework of the APMM. Each maturity level includes a set of statements that cover all six architecture process activities used in this study. The number of statements varies for each maturity level and for each architecture process activity. Throughout the rest of this paper, abbreviations for each key process activity will be used. These include domain engineering (DE), requirements management and modeling (RMM), commonality management (CM), variability management (VM), architecture analysis and evaluation (AAE) and architecture artifact management (AAM). The following sub-sections describe the characteristics of an organization dealing with the SPL. Specifically, organizations will be described in terms of the architecture maturity scale and the measuring instrument designed particularly for this APMM. In describing the measuring instrument, we will use the following symbols and abbreviations:

**APA.X.Y**

APA= Architecture Process Activity
X = Maturity Level (an integer)
Y = Architecture Process Activity Number (an integer)

**S.I.J.K**

S = Statement
I = Maturity Level (an integer)
J = Architecture Process Activity Number (an integer)
K = Statement Number (an integer)

**Table II: Framework of Architecture Process Maturity Model**

| Maturity Level | Architecture Process Activities & Number of Statements in Assessment Questionnaires | | | | | | |
|---|---|---|---|---|---|---|---|
| | DE | RMM | CM | VM | AAE | AAM | Total |
| Independent Product Development | 2 | 5 | 2 | 2 | 2 | 2 | 15 |
| Standardized Infrastructure | 3 | 4 | 3 | 3 | 3 | 3 | 19 |
| Software Platform | 4 | 5 | 2 | 4 | 4 | 3 | 22 |
| Software Product Family | 5 | 3 | 3 | 4 | 3 | 2 | 20 |
| Configurable Product Base | 4 | 3 | 3 | 4 | 3 | 2 | 19 |

## 1) Independent Product Development (Level 1)

The "Independent Product Development" stage of the SPLA process indicates that an organization does not have a set of stable and organized architecture process activities for SPL development. In an organization at this level, there is a lack of understanding about the significance of the SPLA process. Furthermore, there is no evidence that the organization performs SPL engineering activities in a coordinated way. Instead, the organization tends to develop multiple products independently and reuse their software assets on an ad hoc and as-needed basis. Also, there is no defined protocol to switch from a single product to a line of products that share a common architecture. The organization does not have the technological resources and skills to establish a SPLA despite the fact that they have a growing interest in setting up an infrastructure for a SPL. The following measuring instrument illustrates the SPLA maturity of an organization when it is at Level 1 in terms of key architecture process activities.

APA.1.1  Domain Engineering

    S.1.1.1  The organization does not have an established unit to perform domain engineering, and most of the activities are performed solely on an ad hoc and as needed basis.
    S.1.1.2  The organization does not have sufficient knowledge about the domain of SPL.

APA.1.2  Requirements Management & Modeling

    S.1.2.1  SPLA requirements are not clearly defined and identified.
    S.1.2.2  The requirements are managed at individual product levels.
    S.1.2.3  There is a lack of technical understanding regarding SPLA requirements.
    S.1.2.4  The organization is not using any modeling techniques to elaborate SPLA requirements.
    S.1.2.5  The organization does not have the technical means and knowledge to model SPLA requirements.

APA.1.3  Architecture Analysis and Evaluation

    S.1.3.1  The organization lacks an understanding of SPLA analysis techniques.
    S.1.3.2  There is no evidence that the organization performs a systematic analysis of SPLA.

APA.1.4  Commonality Management

    S.1.4.1  The commonality among independent products does not result from any planning.
    S.1.4.2  The commonality among products results from the ad hoc reusability of software assets.

APA.1.5  Variability Management

    S.1.5.1  There is no evidence of planned variability among successive products.
    S.1.5.2  The SPLA does not define any variation points

APA.1.6  Architecture Artifact Management

    S.1.6.1    SPLA artifacts are not maintained and documented.
    S.1.6.2    The requirements are documented only at the individual product level.

## 2) Standardized Infrastructure (Level 2)

The next architectural maturity level of the SPL is Level 2, also known as "Standardized Infrastructure". The organizations at this level aim to develop a SPLA and encourage employees to acquire and share knowledge and skills for SPL engineering. At this earlier stage, the organization is concentrating their efforts on creating a domain engineering unit to initiate the development of an infrastructure for their SPLA. The organization understands the significance of modeling architectural structures and patterns and is currently developing its expertise to manage and model SPLA requirements. Furthermore, the organization understands the importance of commonality and variability management in SPLA, but there is a lack of systematic and planned management of the commonality and variability among products. Also, there are no clear guidelines or methodologies to evaluate the SPLA. The organization is not maintaining the appropriate documentation for their SPLA. Overall, the organization understands the importance of a SPLA and they are in the process of establishing an infrastructure for the SPL. The measuring instrument below illustrates the set of statements that must be satisfied for an organization to achieve Level 2.

APA.2.1  Domain Engineering

    S.2.1.1    The organizational structure clearly defines and supports the presence of a domain-engineering unit.
    S.2.1.2    The roles and responsibilities in the domain-engineering units are not yet explicitly defined.
    S.2.1.3    The organization is acquiring knowledge about the domain of the SPL.

APA.2.2  Requirements Management & Modeling

    S.2.2.1    The organization is making an effort to acquire technical knowledge and to understand the managing of SPLA requirements
    S.2.2.2    The organization collects and analyzes data from the consumer market in order to identify the potential requirements of SPLA.
    S.2.2.3    The organization is using a notation language to model SPLA requirements.
    S.2.2.4    The organization understands that requirement models facilitate the understanding of SPLA requirements, but there is a lack of technical knowledge for developing architectural models.

APA.2.3  Architecture Analysis and Evaluation

    S.2.3.1    The organization is acquiring knowledge and skills to analyze the SPLA.
    S.2.3.2    The organization has not yet established clear guidelines or a well-documented methodology to evaluate the SPLA.
    S.2.3.3    The quality and functional attributes necessary for evaluating the SPLA are not yet defined.

APA.2.4  Commonality Management

    S.2.4.1    The organization understands the importance of commonality among successive products.
    S.2.4.2    There is a lack of systematic and planned management of the commonality among products.
    S.2.4.3    The organization is continuously learning to manage commonality among products and to avoid making mistakes in this endeavor.

APA.2.5  Variability Management

    S.2.5.1    There is a lack of systematic and planned management of the variability among products.
    S.2.5.2    The uncontrolled variability among products is a response to the actions of competitors.
    S.2.5.3    The organization is acquiring knowledge and skills to handle the variability among products.

APA.2.6  Architecture Artifact Management

S.2.6.1 The significant architectural requirements are identified but the organization does not systematically document these requirements.
S.2.6.2 The architectural structure is identified but the organization is not using any architectural description language to document the structure, the sub-units or the connection among them.
S.2.6.3 The component description, interface requirements, interconnection hierarchy and variation mechanisms are not documented.

## 3) Software Platform (Level 3)

An organization at Level 3, or the "Software Platform," is able to establish an infrastructure of SPLA by completing a comprehensive domain engineering activity. The strategic plans show the organizational commitment to developing a SPLA. SPLA requirements are identified and documented as a result of sufficient knowledge about the domain. The organization prepares and manages requirement models, which represent the structural layout and the interconnection among various architectural sub-units. Subsequently, the organization employs the use of architecture description language to document components, interfaces, classes, and objects. The domain engineering activities in the organization identify commonality and variability among a set of envisioned product line applications. Specifically, the commonality and variability among products is explicitly identified in the models of the SPLA. The organization has established clear guidelines and a well-documented methodology to evaluate the SPLA. Accordingly, the employees are trained with the required knowledge of a SPLA methodology. Overall, the organization understands the process methodology of SPLA and is able to streamline activities for SPL engineering from the architectural aspects. The subsequent measuring instrument illustrates the set of statements designed for Level 3.

APA.3.1 Domain Engineering

S.3.1.1 The roles and responsibilities of individuals and groups are well-defined and documented in the organization's domain and engineering units.
S.3.1.2 The domain requirements of the SPL are clearly defined, stated and documented.
S.3.1.3 The organization has sufficient knowledge of the SPL domain.
S.3.1.4 The domain engineering activity for the product line identifies the potential market segment.

APA.3.2 Requirements Management & Modeling

S.3.2.1 The organization has acquired sufficient knowledge and technical ability to manage SPLA requirements.
S.3.2.2 The requirements of the SPLA are clearly identified and well documented.
S.3.2.3 The requirements model explicitly shows the structural layout of the product line architecture.
S.3.2.4 The requirements model envisions the development of product lines.
S.3.2.5 The requirements model helps in visualizing the inter-connection of various architectural sub-units.

APA.3.3 Architecture Analysis and Evaluation

S.3.3.1 The organization has established clear guidelines and a well-documented methodology to evaluate the SPLA.
S.3.3.2 The simulations and prototyping activities are used to analyze the structure of and interconnection among the SPLA components.
S.3.3.3 The organization is using standard industry practices to evaluate SPLA.
S.3.3.4 The organization has acquired sufficient knowledge and technical abilities to evaluate their SPLA.

APA.3.4 Commonality Management

S.3.4.1 The domain engineering activities in the organization identify commonalities among a set of envisioned product line applications.
S.3.4.2 The commonality among products is explicitly identified in the SPLA.

APA.3.5 Variability Management

S.3.5.1 The domain engineering activities in the organization identify variability among a set of envisioned product line applications.

S.3.5.2 The organization identifies the variability among products by showing the areas of variation in the SPLA.
S.3.5.3 The organization documents the variability in components, interfaces, classes, and objects, and their design documents highlight the areas of variation.
S.3.5.4 The variability information is available to the application-engineering unit when necessary.

APA.3.6 Architecture Artifact Management

S.3.6.1 The organization is using an architectural description language to describe and document architectural structure and textures.
S.3.6.2 Significant architectural requirements are well documented and traceable.
S.3.6.3 The architectural layers and design decisions are well documented and traceable.

## 4) Software Product Family (Level 4)

Level 4 of the SPLA maturity is known as the "Software Product Family.. The organization at this level is able to establish a SPLA. Specifically, the scope of the SPLA is clearly defined and documented and it details the product line requirements. The organization develops and manages variability and commonality models to introduce controlled variability and to maximize the commonality among successive products. Furthermore, the organization explicitly defines and utilizes the quality and functional attributes to evaluate the SPLA. The components description, interface requirements, interconnection hierarchy, and variation mechanisms are explicitly documented and traceable. Effective communication channels are present in the organization in order to resolve architecture related issues. The organization is committed to learning and to improving their knowledge in the area of SPLA. The organizational structure supports SPL engineering and there is evidence of strong communication between the domain and application engineering units. Also, the process activities among various departments and sub units are synchronized. The resulting measuring instrument illustrates the set of statements that apply to an organization at Level 4.

APA.4.1 Domain Engineering

S.4.1.1 The SPL scope is well defined and documented as a result of comprehensive domain engineering activities.
S.4.1.2 Domain analysis identifies a potential set of products for the SPL.
S.4.1.3 The domain-engineering unit generates new ideas and innovations and they take the initiative to experiment with new ideas.
S.4.1.4 The domain-engineering unit works in a collaborative way and provides feedback to other units within the organization.
S.4.1.5 Business plans are based on comprehensive domain engineering activities.

APA.4.2 Requirements Management & Modeling

S.4.2.1 The SPLA requirements comprise the scope of the SPL.
S.4.2.2 The organization has an established and defined inter-communication protocol among external and internal entities for analyzing and identifying SPLA requirements.
S.4.2.3 The organization develops and manages variability models to introduce controlled variability among successive products.

APA.4.3 Architecture Analysis and Evaluation

S.4.3.1 The quality and functional attributes that evaluate the SPLA are explicitly defined.
S.4.3.2 The organization has defined specific qualitative metrics to evaluate the performance of the SPLA.
S.4.3.3 The organization is committed to learning and improving their knowledge in the area of SPLA evaluation.

APA.4.4 Commonality Management

S.4.4.1 The management encourages as much commonality as possible and developers concentrate more on product specific issues rather than on issues common to all products.
S.4.4.2 SPL requirements clearly identify, model and document commonality in products.

S.4.4.3   A well-defined organizational unit with a clear set of guidelines handles the management of core SPL assets, which increase the commonality among products.

APA.4.5   Variability Management

S.4.5.1   Variability among products is within the scope of the SPL.
S.4.5.2   Market requirements and customer expectations influence the design decisions for creating variability among products.
S.4.5.3   Requirement models clearly illustrate variability among products by explicitly showing the areas of variation.
S.4.5.4   The variability among products helps to retain current customers.

APA.4.6   Architecture Artifact Management

S.4.6.1   The components description, interface requirements, interconnection hierarchy and variation mechanisms are explicitly documented and traceable.
S.4.6.2   A well-established configuration management system keeps track of all architecture objects.

## 5) Configurable Product Base (Level-5)

The highest architecture maturity level is called the "Configurable Product Base". At this level, the SPLA plays an integral role in the business of the organization. There is strong evidence that various sub-units of the organization work collaboratively to develop and manage the SPLA. Cross-functional teams are established, which oversee the entire SPL process and support the management in decision-making. The organization learns from their experiments in improving the SPLA process methodology and avoids making future mistakes. Hence, learning and acquiring new knowledge about the SPLA is a continuous process in the organization. Domain and application engineering units cooperatively supervise the synchronization of activities in both departments. The SPLA requirements are regularly reviewed and updated when necessary. Furthermore, the effective commonality management allows maximum software reuse in the organization. The organization regularly conducts market reviews and uses customer feedback to manage variability in successive product development. They also support innovation in the SPLA and promote research and development. The organization is continuously improving their process for evaluating the SPLA and experimenting with innovative methods. The following measuring instrument illustrates the set of statements designed for Level 5.

APA.5.1   Domain Engineering

S.5.1.1   The domain engineering unit has access to information from internal and external resources and uses both formal and informal mechanisms to disseminate learning and knowledge within the organization.
S.5.1.2   A joint team from the domain and application engineering units supervise the synchronization of activities in both departments.
S.5.1.3   The business and domain engineering units coordinate the supervision of marketing plans and strategies.
S.5.1.4   The domain-engineering activities support the execution of strategic organizational plans.

APA.5.2   Requirements Management & Modeling

S.5.2.1   The requirements of the SPLA include the targeted market segment.
S.5.2.2   The requirements of the SPLA are regularly reviewed and updated when necessary.
S.5.2.3   The requirements accommodate the quality attributes of the SPLA.

APA.5.3   Architecture Analysis and Evaluation

S.5.3.1   The organization is continuously improving the process of evaluating the SPLA and is experimenting with innovative methods.
S.5.3.2   The roles and responsibilities of individuals and groups in analyzing the SPLA are well-defined and documented.
S.5.3.3   The organization learns from its experience and avoids repeating mistakes in their evaluation of the SPLA.

APA.5.4   Commonality Management

S.5.4.1 The commonality management allows the maximum amount of software reuse in the organization.
S.5.4.2 The organization regularly conducts market reviews and uses customer feedback to update commonalities among successive products.
S.5.4.3 All of the resulting products share a common SPLA.

APA.5.5 Variability Management

S.5.5.1 The organization is continuously improving the process of managing variability among products.
S.5.5.2 The variable requirements of the product line are well defined and documented.
S.5.5.3 The organization regularly conducts market reviews and uses customer feedback to introduce variable features in successive product development.
S.5.5.4 The variability among products helps to retain regular customers and has a tendency to attract new clients.

APA.5.6 Architecture Artifact Management

S.5.6.1 The architectural objects are regularly reviewed, updated, and communicated to the developers.
S.5.6.2 The organization has a well-established change management plan to introduce and manage changes in the architectural objects.

## D. Performance Scale

The maturity level of an organization's SPLA process is determined by measuring their ability to perform key architecture process activities. The qualitative ratings, in descending order, include the statements "Completely Agree (4)", "Largely Agree (3)", "Partially Agree (2)", and "Not Agree (1)". These statements and their numerical value, as described in Table-III, are used to measure each key architecture process activity Overall, these ratings reflect the agreement of the organization with each statement in the questionnaire. The rating of 0 and its corresponding statement, "Doesn't Apply", is designed to increase the flexibility of the model, and it is treated as a rating of 4 in the algorithm. As illustrated in Table III, the performance scales are similar to the BOOTSTRAP methodology [41]. We intentionally based our performance scale on BOOTSTRAP in order to model the architecture dimension of the SPL assessment process on existing popular scales that are already in use and that have been validated and widely accepted in the industry. Consequently, the rating threshold values of the performance scales are also similar to those of BOOTSTRAP. However, we have introduced some changes in the linguistic expressions of the performance scales in order to maintain consistency with the questionnaire design.. Specifically, our questionnaires take a self-assessment approach into account, whereby an organization is able to evaluate the maturity of the SPLA development process by expressing their extent of agreement with the statements.

**Table III: Performance Scale**

| Scale | Linguistic Expression of Performance Scale | Linguistic Expression of BOOTSTRAP | Rating Threshold (%) |
|---|---|---|---|
| 4 | Completely Agree | Completely Satisfied | $\geq 80$ |
| 3 | Largely Agree | Largely Satisfied | 66.7 - 79.9 |
| 2 | Partially Agree | Partially Satisfied | 33.3 - 66.6 |
| 1 | Not Agree | Absent / Poor | $\leq 33.2$ |
| 0 | Doesn't Apply | Doesn't Apply | - |

## E. Rating Method

The rating method adopted in this APMM drives its foundations partially from the BOOTSTRAP algorithm [41] of software process assessment. The structure of the rating method contains different terms such as Performance Rating ($PR_{APA}$), Number of Agreed upon Statements ($NA_{APA}$), Pass Threshold ($PT_{APA}$), Number of Agreed upon Statements of Variability Management ($NA\_VM_{APA}$), Pass Threshold of Variability Management ($PT\_VM_{APA}$) and Architecture Maturity Level (AML). Each of these terms is discussed in detail below.

Let $PR_{APA}$ [I, J] be a rating of the I$^{th}$ architecture process activity at the J$^{th}$ maturity level. Using the performance scale defined in Table III, $PR_{APA}$ [I, J] can be rated in the following manner:

$PR_{APA}$ [I, J] = 4, if the extent of agreement to the statement is at least 80%.
= 3, if the extent of agreement to the statement is between 66.7 and 79.9%.
= 2, if the extent of agreement to the statement is between 33.3 and 66.6%.
= 1, if the extent of agreement to the statement is less than 33.2%.
= 0, if the statement does not apply in this assessment.

An I$^{th}$ statement at the J$^{th}$ maturity level is considered to be agreed upon if $PR_{APA}$ [I, J] $\geq$ 3 or $PR_{APA}$ [I, J] is equal to 0. If the number of statements agreed upon at maturity level "J" is $NA_{APA}$ [J] then it is defined by the following expression:

$NA_{APA}$ [J] = Number of $\{PR_{APA}$ [I, J] | Agreed upon$\}$
= Number of $\{PR_{APA}$ [I, J] | $PR_{APA}$ [I, J] $\geq$ 3 or $PR_{APA}$ [I, J] =0$\}$

Table IV illustrates the pass threshold of 80% at each maturity level. In this table, the values are calculated to the nearest hundred. The maturity level is considered as a pass if 80% of the statements in the questionnaire are agreed upon. If $N_{APA}$ [J] is the total number of statements at the J$^{th}$ maturity level, then the pass threshold ($PT_{APA}$) at the J$^{th}$ maturity level is defined as:

$PT_{APA}$ [J] = $N_{APA}$ [J] * 80%

Variability management is considered as a key characteristic of the product line architecture development process. If the architecture has not been developed with a focus on variability, then there is little chance for successful product line architecture. Consequently, the rating methodology highlights the significance of this factor by assigning importance to variability management in the overall calculation of the architecture process maturity.

If the number of statements agreed upon for "variability management" at maturity level "J" is $NA\_VM_{APA}$ [J] then it is defined by the following expression:

$NA\_VM_{APA}$ [J] = Number of $\{PR_{APA}$ [I = 5, J] | Agreed upon$\}$
= Number of $\{PR_{APA}$ [I = 5, J] | $PR_{APA}$ [I = 5, J] $\geq$ 3 or $PR_{APA}$ [I = 5, J] =0$\}$

If $N\_VM_{APA}$ [J] is the total number of "variability management" statements at the J$^{th}$ maturity level, then the pass threshold ($PT\_VM_{APA}$) at the J$^{th}$ maturity level for "variability management" is defined as:

$PT\_VM_{APA}$ [J] = $N\_VM_{APA}$ [J] * 80%

The Architecture Maturity Level (AML) is defined as the highest maturity level at which the number of statements agreed upon is greater than or equal to the pass threshold ($PT_{APA}$ [J]) and $PT\_VM_{APA}$ [J], defined by:

AML = max $\{J \mid NA_{APA}$ [J] $\geq$ $PT_{APA}$ [J] && $NA\_VM_{APA}$ [J] $\geq$ $PT\_VM_{APA}$ [J]$\}$

**Table IV: Rating Threshold**

| Maturity Level | Total Statements | Overall Pass Threshold 80% | Variability Management Statements | Variability Management Pass Threshold 80% |
|---|---|---|---|---|
| Independent Product Development | 15 | 12 | 2 | 2 |
| Standardized Infrastructure | 19 | 15 | 3 | 2 |
| Software Platform | 22 | 18 | 4 | 3 |
| Software Product Family | 20 | 16 | 4 | 3 |
| Configurable Product Base | 19 | 15 | 4 | 3 |

*All values are calculated to the nearest hundred

# IV. CASE STUDIES

We applied the APMM presented in this work to two organizations currently involved in SPL engineering. In doing so, we performed a maturity assessment on their architecture development process. The two organizations, whose names are kept confidential, agreed to participate in our study. For experimental purposes, the organizations are named "A" and "B". Organization "A" is one of the largest companies in the automobile industry and has been using the concept of SPL engineering in developing embedded systems for various parts of automobiles. Organization "B" is a software development firm that has software development sites worldwide. Table V shows detailed assessment results for Organization "A." The numerical values entered in each cell represent the organization's extent of agreement with the questionnaire statements for each maturity level. Table VI reports the summary of the assessment results. According to the rating method discussed in Section II, Part "D", a statement is considered to be agreed upon if the performance rating shown in Table III is either greater than or equal to 3, or at 0. Thus, based on the results in Table VI, Organization "A" is at the architecture maturity level of "Software Product Family," which is Level 4, whereas Organization "B" is at Level 2 and is considered "Standardized Infrastructure". The following section further highlights the assessment methodology that was used in this study.

## A. Assessment Methodology

- The two participating organizations are from North America. Based on the assumption that a large organization has more than 3000 employees in various departments, these two organizations are considered to be large-scale.
- In the first stage of the study, we established contacts with individuals in the two organizations in order to request their participation in this study. In particular, we sent personal emails to the individuals, stating the scope and objectives of the study. The individuals were working in the area of SPL engineering, which assured us of their competence. Finally, we guaranteed the participants that the assessment conducted for this work was part of a Ph.D. research study and that neither the identity of an individual nor of an organization would be disclosed in the Ph.D. thesis or in any subsequent research publication.
- The questionnaires for each maturity level are designed to serve as a way of learning about the performance of the architectural dimension of SPL engineering. The individuals participating in the study were requested to provide their extent of agreement to each statement by using the performance scale shown in Table III, which ranges from 0 to 4.
- Our assessment methodology uses a top-down approach, where each organization has to start by completing the questionnaire for Level 1 and then progress in increasing order to Level 5. The design of the questionnaire statements are also based on the top-down approach, where more enhanced characteristics are present when moving from a lower to a higher level.
- All of the participants in this study were volunteers and no compensation in any form was offered or paid. We also informed the respondents that they had the option of leaving any statement blank that they did not wish to answer.
- The respondents of this study, on average, had been associated with their respective organization for the last three years. The minimum educational qualification of the respondents was an undergraduate university degree and the maximum was a Ph.D. degree. Most of the respondents belonged to middle or senior technical management and were associated with the software development process. However, some of the participants were from marketing, sales or business development departments. Several of the participants had roles in making policies or implementing organizational strategies from top to bottom.
- We informed the participants of some major sources of data, such as documents, plans, models and actors. This was done in order to reduce the chances of overestimation or underestimation due to poor judgment and to increase the reliability of the approach.
- We did not visit the organizations in person, thus we did not conduct the case studies in the usual way of performing an on-site assessment. Rather, our major source of contact and communication with the participants was email.
- We received more than one response from each organization. Since multiple respondents within one organization may result in conflicting opinions, we performed inter-rater agreement analysis and reported the results in the subsequent section. But receiving more then one response from each organization has also increased the reliability of the assessment methodology by reducing bias up to certain extent.

## B. Inter-rater Agreement Analysis

Since multiple respondents within a single organization may create conflicting opinions about the practice of architectural factors within that organization, we performed and reported inter-rater agreement analysis. Inter-rater agreement corresponds to reproducibility in the evaluation of the same process according to the same evaluation specification [29]. According to El Emam [14] the inter-rater agreement is concerned with the extent of agreement in the ratings given by independent assessors to the same software engineering practices. Thus, in our study, we wanted to discern the extent of agreement among the participants from one organization. In the case of ordinal data, the Kendall coefficient of concordance (*W*) [40] is often preferred to evaluate inter-rater agreement, especially in comparison to other methods such as Cohen's Kappa [10]. "*W*" measures the divergence of the actual agreement shown in the data from perfect agreement. Accordingly, we conducted and reported the inter-rater agreement analysis using Kendall and Kappa statistics. Table VII reports the Kendall and Kappa statistics for Organization "A". Values of Kendall's *W* and the Fleiss Kappa coefficient can range from 0 to 1, with 0 indicating complete disagreement, and 1 indicating perfect agreement [30]. The standard for Kappa [14] includes four level scales: < 0.44 indicates poor agreement, 0.44 to 0.62 means moderate agreement, 0.62 to 0.78 indicates substantial agreement, and > 0.78 entails excellent agreement. In this study, the Kappa coefficient observed ranges from 0.62 to 0.69 and is therefore in the category of substantial.

**Table-V Details of Assessment Result of Case Study "A"**

| Level-1 | | Level-2 | | Level-3 | | Level-4 | | Level-5 | |
|---|---|---|---|---|---|---|---|---|---|
| **Statement #** | **Value** | **Statement #** | **Value** | **Statement #** | **Value** | **Statement #** | **Value** | **Statement #** | **Value** |
| S1.1.1 | 2 | S.2.1.1 | 1 | S.3.1.1 | 4 | S.4.1.1 | 4 | S.5.1.1 | 3 |
| S1.1.2 | 2 | S.2.1.2 | 2 | S.3.1.2 | 4 | S.4.1.2 | 2 | S.5.1.2 | 2 |
| S1.2.1 | 2 | S.2.1.3 | 2 | S.3.1.3 | 4 | S.4.1.3 | 4 | S.5.1.3 | 2 |
| S1.2.2 | 2 | S.2.2.1 | 2 | S.3.1.4 | 4 | S.4.1.4 | 4 | S.5.1.4 | 2 |
| S1.2.3 | 1 | S.2.2.2 | 1 | S.3.2.1 | 3 | S.4.1.5 | 4 | S.5.2.1 | 3 |
| S1.2.4 | 1 | S.2.2.3 | 1 | S.3.2.2 | 3 | S.4.2.1 | 4 | S.5.2.2 | 3 |
| S1.2.5 | 2 | S.2.2.4 | 1 | S.3.2.3 | 4 | S.4.2.2 | 4 | S.5.2.3 | 2 |
| S1.3.1 | 1 | S.2.3.1 | 3 | S.3.2.4 | 4 | S.4.2.3 | 4 | S.5.3.1 | 3 |
| S1.3.2 | 1 | S.2.3.2 | 4 | S.3.2.5 | 4 | S.4.3.1 | 3 | S.5.3.2 | 3 |
| S1.4.1 | 1 | S.2.3.3 | 3 | S.3.3.1 | 4 | S.4.3.2 | 2 | S.5.3.3 | 3 |
| S1.4.2 | 1 | S.2.4.1 | 1 | S.3.3.2 | 4 | S.4.3.3 | 3 | S.5.4.1 | 3 |
| S1.5.1 | 1 | S.2.4.2 | 1 | S.3.3.3 | 4 | S.4.4.1 | 4 | S.5.4.2 | 2 |
| S1.5.2 | 1 | S.2.4.3 | 1 | S.3.3.4 | 4 | S.4.4.2 | 4 | S.5.4.3 | 2 |
| S1.6.1 | 1 | S.2.5.1 | 3 | S.3.4.1 | 4 | S.4.4.3 | 4 | S.5.5.1 | 3 |
| S1.6.2 | 1 | S.2.5.2 | 4 | S.3.4.2 | 4 | S.4.5.1 | 4 | S.5.5.2 | 2 |
| | | S.2.5.3 | 4 | S.3.5.1 | 3 | S.4.5.2 | 3 | S.5.5.3 | 3 |
| | | S.2.6.1 | 4 | S.3.5.2 | 3 | S.4.5.3 | 3 | S.5.5.4 | 3 |
| | | S.2.6.2 | 4 | S.3.5.3 | 4 | S.4.5.4 | 3 | S.5.6.1 | 2 |
| | | S.2.6.3 | 3 | S.3.5.4 | 4 | S.4.6.1 | 2 | S.5.6.2 | 2 |
| | | | | S.3.6.1 | 4 | S.4.6.2 | 3 | | |
| | | | | S.3.6.2 | 4 | | | | |
| | | | | S.3.6.3 | 4 | | | | |

**Table-VI Summary of Assessment Results of Case Studies**

| Maturity Level | Total Statements | Pass Threshold 80% | Organization "A" $NA_{APA}$ | Organization "A" $NA\_VM_{APA}$ | Organization "B" $NA_{APA}$ | Organization "B" $NA\_VM_{APA}$ |
|---|---|---|---|---|---|---|
| Level 1 | 15 | 12 | 0 | 0 | 9 | 0 |
| Level 2 | 19 | 15 | 9 | 3 | 18 | 3 |
| Level 3 | 22 | 18 | 22 | 4 | 10 | 2 |
| Level 4 | 20 | 16 | 17 | 4 | 3 | 2 |
| Level 5 | 19 | 15 | 10 | 3 | 0 | 1 |

$NA_{APA}$ = Total number of agreed upon statements
$NA\_VM_{APA}$ = Total number of agreed upon statements of Variability Management

**Table-VII: Inter-Rater Agreement Analysis**

| Maturity Level | Kendall Statistics | | Kappa Statistics | |
|---|---|---|---|---|
| | Kendall's Coefficient Of Concordance (W) | $\chi^2$ | Fleiss Kappa Coefficient | Z |
| Level 1 | 0.72 | 58.20* | 0.68 | 8.20* |
| Level 2 | 0.65 | 52.90* | 0.63 | 7.98** |
| Level 3 | 0.71 | 57.42* | 0.67 | 8.04* |
| Level 4 | 0.63 | 51.32* | 0.62 | 7.54* |
| Level 5 | 0.74 | 60.14** | 0.69 | 9.01** |

\* Significant at P < 0.01     \*\* Significant at P < 0.05

## C. Limitations of the Assessment Methodology

Certain limitations are implicit in questionnaire based maturity models, of which our study was a part. Some of the limitations associated with this APMM of SPL engineering are as follows:

- Although we used six factors in each of the five maturity levels, there may be other factors that influence the architecture development process, such as organization size as well as economic and political conditions, neither of which are considered in this model.
- Our methodology was limited to subjective assessment. While the statistical techniques that we used to ensure the reliability and validity are commonly used in software engineering, our measurement is still largely based on the subjective assessment of an individual.
- Although we used multiple respondents within the same organization to reduce bias, bias still is a core issue in decision-making. SPL engineering is a relatively new concept in software development, and not many of the organizations in the software industry have institutionalized and launched this concept, so the data collected during this pilot study was limited to a smaller number of individuals, and thus, more susceptible to bias.
- The possibility of participant error and unreliability was present in our study. We asked the respondents to consult major sources of data in their organization, such as documents, plans, models, and actors, before responding to a particular item in order to reduce the human tendency to estimate their organization's activities incorrectly. However, this was largely dependent on individual efforts to collect the required information before responding to the statements in the questionnaire.
- Our assessment methodology does not take the role of the independent assessor into account. In most maturity models, the independent assessors are integral in defining the coordination of the assessor with the internal assessment team performing the evaluation. However, our current case studies are based on self assessment
- The methodology evaluates and provides numerical data about the maturity of the architecture factors and the overall maturity of the architecture dimension, but the maturity model does not provide any guidelines for the improvement process, which we consider as a future work for this study.

Although the APMM presented in this paper has some general and specific limitations, it nevertheless provides a comprehensive approach to evaluating the maturity for the architecture dimension of the SPL engineering process and it provides foundations for future research in this area.

## D. Utilization of the Architecture Process Maturity Assessment Model

One of the advantages of using maturity models in software engineering is to obtain inside information about the current maturity of the different process related activities in an organization. Ideally, this information provides a basis for improvement plans and activities. Furthermore, maturity models are also advantageous to individual organizations because companies with high ratings are attractive to potential customers. We summarized the advantages of the architecture process maturity model from different perspectives such as software engineering research, organizational aspects, product development, and process improvements.

- Overall, the maturity model presented in this work provides information in the form of maturity assessment that can be used to improve the process methodology and complements product development activity in the organization using the concept of software product line engineering.

- The overall performance of an organization depends on the success of managing its core architecture and there are number of critical factors that facilitate the development of product line architecture. Since technologies and business requirements are evolving rapidly, companies must monitor the activities affecting the performance software architecture. The maturity assessment model presented in this work helps companies to monitor and evaluate the activities leads to product line software architecture development in the organization.
- The model presented in this work highlights a methodology to evaluate some of the key architectural process activities in a company. This evaluation provides inside information about the activities that can be improved upon by development team and management. For example, if development team discovered that variability management is at a lower maturity level then they can introduce changes in the variability management protocols to improve it. This improvement will subsequently help in the product development process, which is the ultimate goal of the organization.
- The software product line is gaining popularity and many organizations around the world are currently involved in applying this concept. Our model provides an early conceptual framework for the maturity assessment of software product line engineering. Consequently, this area still requires future contributions from software engineering researchers.

# V. Final Remarks

The engineering efforts for SPL development and management have been divided into the four dimensions of business, architecture, software engineering, and organization. SPL process assessment is an area of immense importance from the perspective of software engineering, especially the SPL. Currently, no work has been done in the area of SPL process maturity assessment apart from a few initial theoretical studies. The conceptual layouts of the SPL process assessment envision the overall methodology as a set of four maturity evaluation frameworks for business, architecture, software engineering, and organization. Subsequently, this research contributes towards establishing a comprehensive and unified strategy for the process assessment of SPL by addressing the architecture dimension. As a result, our work presents an architecture process maturity model for evaluating this dimension of the SPL process methodology. The model provides a methodology to evaluate the current maturity for the architecture dimension of the SPL in an organization. Furthermore, the framework of the model consists of assessment questionnaires for the five maturity levels, performance scales and a rating method. The case studies conducted in this research show the maturity of the SPLA process in two organizations.